\documentstyle[ 12pt]{article}

\begin{document}

\centerline { \bf  VARIABLE FINE STRUCTURE CONSTANT FROM }
\centerline{ \bf MAXIMAL-ACCELERATION PHASE SPACE RELATIVITY}

\bigskip

\centerline { Carlos Castro }

\centerline{ Center for Theoretical Studies of Physical Systems, }

\centerline{ Clark Atlanta University, Atlanta, GA. 30314 }

\centerline { October , 2002 }

\bigskip

\centerline{\bf  Abstract}

\bigskip

We presented a new physical model that links the maximum speed of light
with the minimal Planck scale into a maximal-acceleration Relativity
principle in the spacetime tangent bundle and in phase spaces (cotangent
bundle). The maximal proper-acceleration bound is $a = c^2/ \Lambda$ in
full agreement with the old predictions of Caianiello, the Finslerian
geometry point of view of Brandt and more recent results in the
literature. Inspired by the maximal-acceleration corrections to the Lamb
shifts of one-electron atoms by Lambiase, Papini and Scarpetta, we derive
the $exact$ integral equation that governs the Renormalization-Group-like
scaling dependence of the fractional change of the fine structure constant
as a function of the cosmological redshift factor and a cutoff scale $ L_c
$, where the maximal acceleration relativistic effects are dominant.  A
particular physical model exists dominated entirely by the vacuum energy,
when the cutoff scale is the Planck scale, with $ \Omega_\Lambda \sim 1$ .
The implications of this extreme case scenario are studied.

\bigskip

\centerline {\bf I .  Introduction }

\bigskip

In recent years there has been growing evidence for the Cosmological 
evolution of the fine structure constant using absorption systems in 
the 
spectra of distant
quasars . It was found that the fine structure $ \alpha $ was smaller 
in the 
past . For references see [  11 ] .   The purpose of this note is to 
derive 
an exact integral equation governing the fractional change  $ \Delta 
\over 
\alpha$  of the fine structure due to the Maximal-acceleration 
Relativistic 
effects proposed in [ 1 ] .   The latter were just a consequence of the 
Extended Scale Relativty theory in C-spaces.

Relativity in C-spaces (Clifford manifolds) [1] is a very natural 
extension
of Einstein's relativity and Nottale's scale relativity [2] where the
impassible speed of light and the minimum Planck scale are the two 
universal
invariants.  An event in C-space is represented by a   $polyvector$ , 
or
Clifford-aggregate of lines, areas, volumes, ...... which  bear a 
one-to-one
correspondence to the holographic shadows/projections (onto the 
embedding
spacetime coordinate planes) of a nested family of $p$-loops (closed 
$p$-
branes of spherical topology) of various dimensionalities: $p = 0$ 
represents
a point; $p = 1$ a closed string, $p = 2$ a closed membrane, etc.... 
where
$ p = 0, 1, 2, ....  D-1$.

The invariant ``line"  element associated with a polyparticle is:

$$ d\Sigma ^2 = dX.dX =  ( d \Omega) ^2 + \Lambda ^{2D-2} ( dx_{\mu}dx^
{\mu} ) +
\Lambda ^{ 2D -4 } ( d x_{\mu\nu} ) ( d x^{\mu\nu} ) + ...\eqno(1.1) $$
the Planck scale appears as a natural quantity in order to match units 
and 
combine p-branes ( p-loops ) of different dimensions.   The 
polyparticle 
lives in a target background of  $ D? = D + 1 = p+2 $  dimensions due 
to the 
fact that C-space has $two$ times , the coordinate time $ x_o = t $ and 
the 
$ \Omega$ temporal variable representing the proper $ p+1$-volume.  The 
fact 
that the Planck scale is a minimum
was based on the real-valued interval $ dX $ when $ dX.dX > 0 $. The 
analog 
of photons in C-space are $tensionless$ branes : $ dX.dX = 0 $. Scales 
smaller than
$\Lambda$ yield " tachyonic " intervals $ dX. dX < 0 $ [1]. Due
to the matrix representation of the gamma matrices and the cyclic trace
property, it can be easily seen why the line element is invariant under 
the 
C-
space Lorentz $group$  transformations:
$$Trace ~ X'^{2 } = Trace~ [ R X^2 R^{ - 1 } ] = Trace~ [ R R^{ -1 } 
X^2 ] =
Trace~X^2 ~, \eqno( 1.2) $$
where a finite polydimensional rotation that reshuffles dimensions is
characterized by the C-space ``rotation" matrix:

$$  R = \exp [ i (\theta I + \theta ^{\mu} \gamma _{\mu} + 
\theta^{\mu\nu}
\gamma _{\mu \nu} + ...)] .  \eqno(1.3) $$
The parameters $\theta, \theta ^{\mu}, \theta ^{\mu\nu}, ......$ are 
the 
C-space
extension of the Lorentz boost parameters and for this reason the naïve
Lorentz transformations of spacetime are modified to be:

$$ x'^{\mu} = L^{\mu}_{\nu} ~ [ \theta , \theta ^\mu, \theta^{\mu\nu} , 
...]
x^{\nu}~ +  L^\mu_{\nu \rho } ~ [ \theta, \theta^\mu , \theta^{\mu\nu} 
, 
...] x^{\nu \rho } + .....  \eqno (1.4 ) $$

It was argued in [1] that the extended Relativity principle in C-space 
may
contain the clues to unravel the physical foundations of string and 
M-theory
since the dynamics in C-spaces encompass in one stroke the dynamics of 
all 
p-
branes of various dimensionalities. In particular, how to formulate a 
master
action that encodes the collective dynamics of all extended objects.

For further details about these issues we refer to  [1 ] and all the
references therein. Like the derivation of the minimal length/time
string/brane uncertainty relations;  the logarithmic corrections to the 
black-
hole area-entropy relation;  the existence of a maximal Planck 
temperature ; 
  the origins of a higher derivative gravity with
torsion;  why quantum-spacetime may be  truly infinite dimensional 
whose 
average dimension today is close to $ 4 + \phi^3  = 4.236...$ where $ 
\phi  
= 0.618...$ is the Golden Mean ;  the construction of the p-brane 
propagator;  the role of
supersymmetry;  the emergence of two times; the reason behind a running
value for $ \hbar $ ;   the way to correctly pose the cosmological 
constant 
problem as well as other results.

In  [ 1  ]  we discussed another physical model that links the maximum
speed of light , and the minimal Planck scale,  into a 
maximal-acceleration
principle in the spacetime tangent bundle, and consequently, in the 
phase
space (cotangent bundle). Crucial in order to establish this link was 
the 
use
of Clifford  algebras in phase spaces. The maximal proper acceleration 
bound
is  $a = c^2/ \Lambda$  in full agreement with [4] and the Finslerian
geometry point of view in [6].  A  series of reasons why  C-space 
Relativity 
is more physically appealing  than all the others proposals based
on kappa-deformed Poincare algebras [ 10 ] and other quantum algebras 
was 
presented .

On the other hand, we argued why the truly $bicovariant$ quantum 
algebras 
based on inhomogeneous quantum groups developed by Castellani
[ 15  ]  had a very interesting feature related to the $T$-duality in 
string 
theory ;  the deformation $ q$ parameter could be written as :  $ q = 
exp [ 
\Lambda/ L ] $  and, consequently, the classical limit  $ q = 1 $ is 
attained when the Planck scale $ \Lambda$ is set to zero, but also when 
the 
upper impassible scale $ L $ goes to infinity !  This entails that 
there 
could be $two$ dual quantum gravitational theories with the same 
classical 
limit !  Nottale has also postulated that if there is a minimum Planck 
scale, by duality, there should be another upper impassaible upper 
scale $ L 
$ in Nature [ 2 ] .
For a recent discussion on maximal-acceleration and kappa-deformed 
Poincare  
algebras see  [ 10 ] .  It was  also argued in [ 1 ]  why the theories 
based 
on kappa-deformed Poincare algebras may in fact be related to a Moyal 
star-product deformation of a classical Lorentz algebra whose 
deformation 
parameter is precisely the Planck scale $\Lambda = 1/ \kappa$.

In section {\bf II},  we  discuss  the work in [ 1 ]  and show how to 
derive 
the Nesterenko action [5] associated with a sub-maximally accelerated 
particle in spacetime directly from phase-space Clifford algebras and 
present a  full-fledged C-phase-space generalization of the Nesterenko 
action .

In section { \bf 3 } we review the maximal-acceleration relativistic 
corrections to the Lamb shift of one electron atoms [ 12 ]  ( of 
fractional 
order of $ 10^{ - 5 } $ )
as a preamble to section {\bf 4 } where we derive the exact integral 
equation that governs the Renormalization-Group-like scaling dependence 
of 
the fractional change of the fine structure constant as a function of 
the 
cosmological redshift  factor and a cutoff scale $ L_c $ where the 
maximal 
acceleration relativistic effects are dominant.  To conclude,  we 
derive the 
corrections to the electric charge due to a running Planck constant 
induced 
by Extended Scale Relativistic effects in C-spaces and discuss the 
Cosmological model with $ \Omega_\Lambda  \sim  1 $.

  \bigskip

\centerline { \bf  II.  Maximal-Acceleration from Clifford algebras }

  \bigskip

We will follow closely the procedure described in the book [3] to 
construct
the phase space Clifford algebra. For simplicity we shall begin with a 
two--
dimensional phase space, with one coordinate and one momentum variable 
and
afterwards we will generalize the construction  to higher dimensions.

Let $ e_{p}   e_{q} $ be the Clifford basis elements in a 
two--dimensional
phase space obeying the following relations:

$$  e_p . e_q \equiv  { 1 \over 2 } ( e_q e_p + e_p e_q ) = 0. ~~~ e_p 
. e_p
= e_q . e_q = 1 . \eqno(2.1) $$

The Clifford product of  $ e_p, e_q $ is by definition the sum of the 
scalar
product and wedge product furnishing the unit $bivector$:

  $$ e_p e_q \equiv e_p . e_q +  e_p \wedge e_q  = e_p \wedge e_q =  j  
. 
~~~
j^ 2 =  e_p e_q  e_p e_q = - 1 . \eqno (2.2 ) $$
due to the fact that $ e_p, e_q $ anticommute, eq.~( 2.1).

In this fashion, using Clifford algebras one can justify the origins of
complex numbers without introducing them ad-hoc. The imaginary unit $ j 
$ is
$ e_p e_q $.  For example, a Clifford vector in phase space can be 
expanded,
setting aside for the moment the issue of units, as:

$$  Q = q e_q  + p e_p . ~~~ Q e_q = q + p e_p e_ q = q + j p = z. ~~~
e_q Q = q + p e_q e_p = q - j p = z^*~, \eqno ( 2.3 ) $$
which reminds us of the creation/annhilation operators used in the 
harmonic
oscillator case and in coherent states.

The analog of the action for a massive particle is obtained by taking 
the
scalar product:
$$ dQ . dQ = (dq)^2 + ( dp)^2  \Rightarrow S = m \int  \sqrt { dQ. dQ } 
=
m \int  \sqrt {( dq)^2 + (dp)^2 } . \eqno ( 2.4 ) $$

One may insert now the appropriate length and mass parameters in order 
to
have consistent units:

$$ S = m \int  \sqrt {  ( dq)^2 + ( {  \Lambda \over m  } )^2  (dp)^2 
}.
\eqno (2.5 ) $$
where we have introduced  the Planck scale $\Lambda$  and the mass $m$ 
of 
the
particle to have consistent units, $\hbar = c = 1$.
The reason will become clear below.

Extending this two-dimensional action to a higher $2n$-dimensional 
phase
space requires to have
$ e_{p_{\mu }} ,  e_{q_{\mu}} $ for the Clifford basis where $\mu = 1, 
2,
3...n$. The action in this $2n$-dimensional phase space is:
$$ S = m \int  \sqrt {  ( dq^{\mu} dq_{\mu} ) + ({\Lambda \over m  } 
)^2  
(dp^
{\mu} dp_{\mu})} =
m \int  d \tau \sqrt {1  + ( {  \Lambda \over m  } )^2  (dp^{\mu}/ d 
\tau )
(  dp_{\mu}/ d \tau  ) }
\eqno (2.6 ) $$
in units of $c = 1$, one has the usual infinitesimal proper time
displacement  $d \tau ^2 = dq^{\mu}  dq_{\mu}$.

One can easily recognize that this action (2.6), up to a numerical 
factor of
$ m/a$,  is nothing but the action for a sub-maximally accelerated 
particle
given by
Nesterenko  [5].  It is sufficient to rewrite: $dp^\mu / d \tau  = m 
d^2
x^\mu / d \tau ^2 $  to get from eq.~(2.6):
$$ S = m \int  d \tau \sqrt {  1  + \Lambda^2  (d^2 x^ \mu/ d \tau^2  ) 
(
d^2 x_\mu/ d \tau^2) } . \eqno ( 2.7) $$

Using the postulate that the maximal-proper acceleration is given in a
consistent manner with the minimal length principle (in units of $c = 
1$):

$$ a = c^2 / \Lambda  = 1/\Lambda \Rightarrow
S = m \int  d \tau \sqrt {  1  +   (  { 1 \over a }  )^2  (d^2 x^ \mu/ 
d
\tau^2  ) (  d^2 x_\mu/ d \tau^2   ) }~, \eqno ( 2.8) $$
which is exactly the action of [5], up to a numerical factor of $ m/a 
$,
when  the metric signature is $ ( +, -, - , - ) $.

The proper acceleration is $orthogonal$ to the proper velocity as a 
result 
of
differentiating the timelike proper velocity  squared:

$$ V^2 = {dx^{\mu} \over d \tau } { d x_{\mu} \over d \tau } = 1 =   
V^\mu
V_\mu > 0 \Rightarrow  { d V^{\mu} \over d \tau } V_\mu =  { d^2 
x^{\mu}
\over d \tau ^2 }
V_\mu =  0~, \eqno(2.9)  $$
which means that if the proper velocity is timelike the proper 
acceleration
is spacelike so that:

  $$  g^2 ( \tau ) \equiv -  (d^2 x^ \mu/ d \tau^2  ) (  d^2 x_\mu/ d 
\tau^2)
 > 0  \Rightarrow
S = m \int  d \tau \sqrt {  1  -    { g^2  \over a^2  }     } \equiv  m 
\int
d \omega ~, \eqno(2.10 ) $$
where we have defined:

$$  d \omega  \equiv \sqrt {  1  -    { g^2  \over a^2  }     } d \tau 
. 
\eqno
(2.11) $$
The dynamics of a submaximally accelerated particle in Minkowski 
spacetime
can be reinterpreted as that of a particle moving in the spacetime 
$tangent-
bundle$ background whose $Finslerian$-like metric is:

$$  d\omega^2 = g_{\mu \nu} ( x^\mu, dx^\mu ) dx^\mu dx^\nu  =
(d \tau)^2  (  1  -    { g^2  \over a^2  }     )                . 
\eqno(2.12)
$$

For uniformly accelerated motion, $ g ( \tau ) = g = constant$ the 
factor:

$$ { 1 \over \sqrt {  1  -  { g^2  \over a^2  }  }   }  \eqno (2.13) $$
plays a similar role as the standard Lorentz time dilation factor in
Minkowski spacetime.

The action is real valued if, and only if, $ g^2  < a^2 $ in the same 
way
that the action in Minkowski spacetime is real valued if, and only if, 
$v^2 
<
c^2$. This explains why the particle dynamics has a bound on proper-
accelerations. Hence, for the particular case of a $uniformly$ 
accelerated
particle whose trajectory in Minkowski spacetime is a hyperbola, one 
has an
Extended  Relativity of $uniformly$ accelerated observers whose proper-
acceleration have an upper bound given by  $c^2/ \Lambda$. Rigorously
speaking, the spacetime trajectory is obtained by a canonical 
projection of
the spacetime tangent bundle onto spacetime. The invariant time, under 
the
pseudo-complex extension of the Lorentz group [8], measured in the 
spacetime
tangent bundle is no longer the same as $\tau$,  but instead, it is 
given by
$\omega ( \tau )$.

This is similar to what happens in C-spaces, the truly invariant 
evolution
parameter is not
$\tau$ nor $\Omega$, the Stuckelberg parameter [3], but it is $\Sigma$  
which
is the world interval in C-space and that has units of $length ^ D$. 
The
$group$ of C-space Lorentz transformations preserve the norms of the
Polyvectors and these have units of hypervolumes; hence C-space Lorentz
transformations are volume-preserving.

Another approach to obtain the action for a sub-maximally accelerated
particle was given by [8] based on a pseudo-complexification of 
Minkowski
spacetime and the Lorentz group that describes the physics of the 
spacetime
tangent bundle. This picture is not very different form the Finslerian
spacetime tangent bundle point of view of Brandt [6].  The invariant 
group 
is
given by a pseudo-complex extension of the Lorentz group acting on the
extended coordinates $ X = a x^\mu + I v^\mu $ with $ I^2 =  1 $ 
(pseudo-
imaginary unit) where both position and velocities are unified on equal
footing. The invariant line interval is
$a^2 d \omega^2 = (dX)^2$.

A C-phase-space generalization of these actions (for sub-maximally
accelerated particles, maximum tidal forces) follows very naturally by 
using
polyvectors:
$$ Y =  q^\mu e_{ q_\mu  } +   q^ { \mu \nu } e_{ q_\mu  } \wedge e_{
q_\nu  } +
q^ { \mu \nu\rho } e_{ q_\mu  } \wedge e_{ q_\nu  } \wedge e_{q_\rho} + 
....$$
$$ + p^\mu e_{ p_\mu  } +   p^ { \mu \nu } e_{ p_\mu  } \wedge e_{ 
p_\nu  }
+ ...~,\eqno (2.14) $$
where one has to insert suitable powers of $\Lambda$ and $m$ in the 
expansion
to match units.

The C-phase-space action reads then:

$$S \sim \int \sqrt { dY . dY } = \int \sqrt {   dq^\mu dq_\mu +  d 
q^{\mu
\nu } dq_{\mu \nu } + ... + dp^\mu dp_\mu +  d p^{\mu \nu } dp_{\mu \nu 
}
+ .....}~.
\eqno (2.15 ) $$

This action is the C-phase-space extension of the action of Nesterenko  
and
involves quadratic derivatives in C-spaces which from the spacetime
perspective are effective {\em higher} derivatives theories [1 ]  where 
it
was shown why the scalar curvature in C-spaces is equivalent to a  
higher
derivative gravity. One should expect a similar behaviour for the 
extrinsic
curvature of a polyparticle motion in C-spaces. This would be the 
C-space
version of the action for rigid particles [7]. Higher derivatives are 
the
hallmark of {\em W}-geometry (higher conformal spins).

Born-Infeld models  have been connected to  maximal-acceleration [8]. 
Such
models admits an straightforwad formulation using the geometric 
calculus of
Clifford algebras. In particular one can rewrite all the nonlinear 
equations
of motion in precise Clifford form [9]. This lead that author to 
propose the
$nonlinear$ extension of Dirac's equation for massless particles due to 
the
fact that spinors are nothing but right/left ideals of the Clifford 
algebra:
i.e., columns, for example, of the  Maxwell-Field strength bivector
$F = F_{\mu\nu} \gamma ^{\mu} \wedge \gamma ^{\nu}$.

\bigskip

\centerline{ \bf  III . Maximal-Acceleration corrections to the 
Lamb-shift }

\bigskip

The maximal-acceleration corrections to the Lamb-shift of one electron 
atoms 
were calculated by  Lambiase et al in [ 12 ] .  They started from the 
Dirac 
equation and splitted the spinor into a large and small component. The 
crucial point  behind this calculation was based on the fact that the 
spacetime metric is only flat up to a conformal factor which depends on 
the 
acceleration :

$$ { \tilde g}_{\mu\nu}  =  g_{\mu \nu}  \sqrt  {  1   -  { g^2 ( \tau 
) 
\over A_{max}^2   }  }   .
\eqno ( 3.1 ) $$
This is a Finslerian-type metric.

As a result , the Dirac equation is modified accordingly due to the 
conformally-scaled vierbiens $e^a_\mu  ( x ) =  \sigma ( x ) 
\delta^a_\mu $ 
used to write down the Dirac equation in terms of the ( spacetime ) 
gamma 
matrices $ \gamma^\mu ( x ) = e^\mu_a ( x ) \gamma^a $ :

$$  [  i \hbar \gamma^a  ( \partial_a  +   i { e \over \hbar c } A_a )  
+
i { 3 \hbar \over 2 \sigma } \gamma^a  ( \partial_a \sigma ) - m ] \psi 
( x 
) = 0 . \eqno ( 3.2 ) $$
As usual, the tangent spacetime indices are represented by $ a, b  = 
1,2,3,4 
$.
and one introduces a minimal EM coupling . The Dirac Hamiltonian is 
modified 
and includes the small acceleration-dependent perturbation  :

$$ \Delta H =  - i { 3 \hbar c \over 2 } \gamma^0 \gamma^a  \partial_a 
( - 
\sigma^{ -1} ) . . \eqno ( 3.3) $$

The conformal factor can be expressed in terms of the 
maximal-acceleration 
and a cutoff scale $ r_o $ as follows :

$$ \sigma ( x ) = \sqrt {  1  - { g^2 \over A^2 } }  =    \sqrt {  1  - 
( { 
r_o \over r } )^4  }.
\eqno (  3.4 ) $$

where :

$$ r_o^2  =  {  e^2 \over m A } \sim  3.3 \times 10^{ 16 } cm . \eqno ( 
3.5 
) $$

The corrections to the Lamb-shift were obtained via perturbation theory 
by 
spiltting the spinors into a large/small parts. The corrections to the 
energy spectrum  were :

$$ \Delta E =  < nljm |  \Delta H | nljm > . \eqno ( 3.6 )  $$

In the special case of non-relativistic electrons in an electrostatic 
field 
the maximal acceleration corrections for a one-electron atom  were  :

$$ \Delta E =  6 K  \int  d^3 r  \phi^{+} { 1 \over r^6 }  \phi  -
4K   \int  d^3 r  \phi^{+} { \partial \over r^5  \partial r }  \phi . 
\eqno 
( 3.7 ) $$

with

$$ K \equiv  { 3 \hbar^2 4 m }  ( { e^2 \over m A } )^2 . \eqno ( 3.8 ) 
$$
that has units of $ energy \times length^6 $.

The quantity  defined in terms of the Bohr radius  $ a_o $ has energy 
units  
:

$$  { K \over a_o^6 } = 1.03 ~ kHz . \eqno ( 3.9 ) $$

All the corrections to the Lamb shift  were of the form  :
$$  { K \over a_o^6 } F ( { a_o \over \Lambda } )  e^{ - \Lambda/ a_o } 
. 
\eqno ( 3.10) $$

where $ \Lambda  \leq a_o $ was a suitable cutoff  and the function $ F 
$ 
was a polynomial one.  Lamb-shifts  of the states $ 2p$  and $ 1s$ were 
calculated for a series of values of the $ a_o/\Lambda $ ratios.

( i ) the $ \Delta E^{ ( 2, 0 )} - \Delta E^{ ( 2, 1 ) }  $  and

( ii )  the   $ \Delta E^{ ( 1, 0 )} - \Delta E^{ ( 2, 0 ) }  $  
values.

The most salient feature is that all the maximal-acceleration 
corrections to 
the Lamb-shifts were found to be $positive $ . For example,  taking  
the 
cutoff $ \Lambda $ to be  $ a_o/2.5$, the Lamb-shifts were found to be  
$  
6.9 ~ kHz $ and $ 50.95~kHz$ respectively. Comparing these shifts with 
the 
experimental values of $ 1057851 (4) ~ kHz $ and $ 8172874 ( 60 ) ~kHZ$  
one 
obtains the positive-valued ratios

$$  { 6.9 \over 1057851 } = 0.65 \times 10^{ -5 } . ~~~ { 50.95 \over 
8172874 } = 0.62 \times 10^{ -5 } . \eqno ( 3.11 ) $$

Therefore, the fractional corrections to the fine structure constants 
were 
of the order of $ 10^{ -5 } $. Notice that these correctiosn are 
$positive$-valued and $coincidentally$ agree, in orders of magnitude,  
with 
the observed changes in Astrophysical observations of the fine 
structure 
constant in quasar sources
$  - 0.7 \times  10^{-5} $  [ 11 ] .

The main purpose  of this review  of [ 12 ] is to indicate how, 
coincidentally,  the maximal-acceleration corrections to Lamb-shift of 
one 
electron atoms yield fractional changes to the fine structure constant 
of 
the same order of magnitude as those provided by Astrophysical 
observations 
[ 11 ] ,  with the main difference that there is a positive  variation  
in [ 
12 ] .   A Planck scale cutoff yields lower corrections to the Lamb 
shift as 
expected.  In this case the cutoff $ \Lambda$ cannot be of the order of 
th 
Bohr radius
[ 12 ] .   In addition, at Planck scales EM has to be replaced by a 
more 
fundamental theory.

\bigskip

\centerline{ \bf IV  Variable-Fine Structure in Astrophysics  from 
Maximal-Acceleration }
\bigskip
To explain and derive the observed values of the variable fine 
structure 
constant in Astrophysics from this new physical model based on a 
Maximal-acceleration Phase Space Relativity  principle, we must rely on 
the 
recent observations that the
Universe expansion is in fact accelerating,  contrary to past 
expectations,

To find the the maximal-acceleration corrections to the observed 
variable 
fine structure in Astrophysics we must recall the 
maximal-proper-acceleration Lorentz-like dilation factors  of sections 
{\bf 
2, 3 } :

$$  d\omega   =   d\tau  \sqrt {  1  -   { (d^2 x^\mu / d\tau^2)   (d^2 
x_\mu / d \tau^2) } \over{  A^2 }   }  . \eqno ( 4.1 ) $$

we interpreted in {\bf 2,3   } the  dilation/contraction factors 
stemming 
from the conformal factor associated with the effective Finslerian-like 
metric  :

$$ { \tilde g}_{\mu\nu}  =  g_{\mu \nu}  \sqrt  {  1   -  { g^2 ( \tau 
) 
\over A^2   }  }   .
\eqno ( 4.2 ) $$

Next,  we must find the cosmological dilation/contraction factor which 
is 
responsible for the variation of the electric charge  over cosmological 
time 
( redshifts ) .
The crux is to find the explicit relation between the scaling factor in 
eq-( 
4-1) and the cosmological  redshift factor $ 1+ z  = a (t_o ) /a (t_1 )  
$, 
given by the ratios of the Universe sizes in the present  and past 
epoch . 
The past and present
hypersphere ( hyper-pseudopshere ) radius $ a ( t_1 ) R,  a(t_o)  R $
associated with the Robertson-Walker line element  are given by :

$$  ds^2  = dt^2  - a^2 ( t ) [  { dr^2  \over  1 - k   ( r^2 / R^2 )  
} +
r^2 (  d\theta^2 + sin^2 \theta d \phi^2 )  ] .  \eqno ( 4-3 ) $$
where the parameter $ k $ takes the values $ 0, -1 $ for an open and 
noncompact universe and  $ k = + 1 $ for a closed compact one.   The 
size of 
the universe at any moment of time is represented by  the scaling 
factor $ 
a( t ) $  times the  characteristic
length parameter $ R $.  The cosmological equations for the expansion 
rates 
are obtained after using  Einstein  gravitational equations  [ 13 ] :

$$  {  ( d^2 a/ dt^2) \over a } =  { \Lambda \over 3 }   -  { 4 \over 3 
} 
\pi G ( \rho + 3 p ) $$

$$  (    { ( d a / dt ) \over a }   )^2  =   { 8 \over 3 } \pi G \rho  
\pm  
{ 1 \over a^2 R^2 }  +
{ \Lambda \over 3 } . \eqno ( 4.4 ) $$

The $ \pm$ signs  in  front of the  $ R$ terms in eq- (4-4  ) 
correspond to 
an  open  or closed Universe, respectively.    If the mean mass density 
is 
dominated by nonrelativistic matter  [  13 ]  , then
$ p < < \rho \sim   \rho_o (  { a ( t_o ) \over a  ( t )}  )  ^3   $  .
The redshift factor is  defined as :
$$ 1 + z  =  a ( t_o ) / a ( t ) . \eqno ( 4.5 ) $$
where $ a ( t_o ) $ is the present scale size  and $ a ( t ) $ is the 
past 
scale size of the Universe.

Eqs-(4-4) become :

$$ {  ( d^2 a/ dt^2) \over a }   =
H_o^2  [   \Omega_\Lambda  -  { \Omega_m \over 2 }  ( 1 + z )^3    ]  
$$

$$ (    { ( d a / dt ) \over a }   ) =
H _o  [  \Omega_m ( 1 + z )^{3}  \pm  \Omega_R  ( 1 + z )^{2} + 
\Omega_\Lambda ]^{ 1/2 } . \eqno ( 4.6 ) $$
the fractional contributions to the present value of the Hubble 
constant  $ 
H_o$
due to the mass density,  the radius of curvature and the cosmological 
constant are respectively :

$$  \Omega_m =  { 8 \pi G \rho_o \over 3 H^2_o } . ~~~
\Omega_R  =   {  1 \over  ( H_o a_o R )^2  } . ~~~ a_o R = R_o = 
R(t_o). ~~~
\Omega_\Lambda =  { \Lambda \over 3 H^2_o} . \eqno ( 4.7 )  $$

with

$$ \Omega_m \pm \Omega_R  + \Omega_\Lambda  = 1 $$
( for an open/closed  Universe  respectively ).

Rigorously speaking we should take derivatives w.r.t the  proper-time 
rather 
than the cosmological time $ t $ and instead of solving the Einstein 
equations one should be solving the full-fledged C-space Gravitational 
equations which are equivalent to a Higher Derivative Gravity with 
Torsion. 
This is a very difficut task. For this reason to get an estimate of the 
maximal-acceleration corrections to the electric charge as a result
of the cosmological expansion we shall study the simplest model above.

Inserting the values   $ R ( t ) = a ( t ) R $  into eqs- (  4-1, 4-2    
) 
yields  after some straightforward algebra  the relativistic factor 
analog 
of $ \beta^2  = (v/c)^2  $ :

$$ \beta^2 =  F(z)^2  =  (   { d^2 R(t)/ dt^2   \over  A_{max}  } )^2  
=
  ( H_o  R(  t  ) )^2  ( H_o L)^2
[   \Omega_\Lambda  - { \Omega_m \over 2 }  ( 1 + z )^3   ] ^2 . \eqno 
( 4.8 
)    $$
where we  define the maximal-acceleration in terms of a  length cuttoff  
$ 
L_c $  in units  of $c   = 1 $  :

$$  A_{max }  =  { c^2 \over L_c } = { 1\over L_c  }. \eqno ( 4.9 )  $$

such as :

$$  L_{Planck } <  L_c   <    R ( t_{past} )   < R ( t_{today } ) . 
\eqno ( 
4.10) $$

Let us suppose that we wrote the ordinary electrostatic Coulomb energy  
:

$$ e^2/ L  . \eqno ( 4.11 ) $$
with the fundamental difference now that the length  interval $ L$   is 
no 
longer equal to the flat space one ,  but  it changes as a result of an 
effective metric that is now conformally flat .  The scaling factor is 
dependent on the acceleration    :

$$  d L  =   c d \tau \sqrt {  1  -  { g^2 ( \tau )} \over{ A^2 } }  . 
\eqno ( 4.12 ) $$

For a  $small$  and uniform acceleration $ g ( \tau ) = g  = constant 
$, we  
may expand the square root in a powers series  and integrate  to get :

$$ L \sim  c \tau   (  1   -  {  g^2 \over  2  A^2 }   ) . \eqno ( 
4.13)  $$

$$  L/ c \tau  \sim    1   -   {  g^2 \over  2  A^2 } \Rightarrow
{ L - c \tau \over c \tau }  \sim      -    {  g^2 \over  2  A^2 }  . 
\eqno 
( 4.14 ) $$

Hence, in the small and uniform acceleration limit we have  fractional 
length
changes:

$$  { \Delta  L \over L }   =  -   {  g^2 \over  2  A^2 }  . \eqno ( 
4.15) 
$$

When the maximal proper acceleration is taken to $\infty$ the 
fractional 
length change is $0$.  Since the proper lengths are now scaled by a 
conformal factor dependent on the acceleration,   one could $reabsorb$ 
such 
scalings by a suitable scaling of the electric charges , which in turn, 
will 
modify  the fine structure constant :

$$     { \Delta  \alpha \over \alpha}  =   -  { g^2   \over  {2 A^2 } } 
$$

And in magnitude we may write :

$$  |  { \Delta  \alpha \over \alpha}  |   =    { g^2   \over { 2 A^2 } 
} . 
\eqno ( 4.16) $$ .

Therefore, we propose that the physical mechanism responsible for a 
variation of the fine structure constant is  due to the 
Maximal-acceleration 
Relativity principle ! .  This is the most important conclusion of this 
work.
We have presented a very specific example for a small and uniform 
acceleration that permits  us to expand in a power series and to 
integrate 
trivially . In general this is not the case.

The  more general expression  is  :

  $$  { \Delta  \alpha \over \alpha}   =    - 1 +
   {  c \over ( L_1 - L_o )  }  { 1 \over   \sqrt  { 1 -  
g^2(\tau_o)/A^2 }  
}
\int _{\tau_o}^{ \tau_1} ~  d \tau   \sqrt {  1  -  { g^2 ( \tau )  
\over 
A^2 }    }   . \eqno ( 4.17)   $$

When the maximal acceleration $ A = \infty $ one gets :

$$   { \Delta  \alpha \over \alpha}  =   { L_1 - L_o \over L_1 - L_o } 
- 1 = 
0. \eqno ( 4.18)  $$
as it should.  We shall now use these results to obtain the precise 
expression for the variations of the fine structure constant within the 
Cosmological scenario of a Robertson-Walker-Friedmann model with a 
nonvanishing cosmological constant.

The fundamental equation that governs the cosmological evloution of the 
fine 
structure constant  as function of the redshift  is  then :

$$  { \Delta ( \alpha )  \over   \alpha  } =  - 1  +
  { 1 \over  ( t(z_o)  - t ( z_1) ) } { 1 \over   \sqrt { 1 - F^2 ( 
z_o) }   
}
\int _{z_o}^{z_1}  dz   ~ (d t  ( z ) / dz )   ~   \sqrt  { 1 - F^2 ( z 
) }  
      . \eqno ( 4.19 )  $$
where  the time integration  is taken from the present  $ t_o $ to the 
past 
$ t_1 $ ,  which can be converted into an integration over the redhsift 
factor from the present
$ z_o $  to the past   $ z_1 $  by noticing  that  $ dt >  0 
\leftrightarrow 
  dz < 0 $ :
$$  d t ( z )    =   {  d t ( z ) \over d z }   d z    =
{ dz \over H_o  ( 1 + z )  E ( z )  }  \equiv   { dz \over H ( z )} . 
\eqno 
( 4.20 )  $$
where  [ 13 ] :
$$ E ( z )  \equiv
[ \Omega_m ( 1 + z )^3  \pm  \Omega_R  ( 1 + z )^2 + \Omega_\Lambda ]^{ 
1/2 
} $$
The redshift dependence of the Hubble parameter is :

$$ H ( z ) \equiv H_o ( 1 + z ) E ( z ). \eqno ( 4.21 )  $$
naturally :
$$  E ( z = 0 ) = 1 = [  \Omega_m  \pm  \Omega_R   + \Omega_\Lambda ]^{ 
1/2 
} $$
The net fractional contribution  to the mass-energy density from the 
three 
sources has to add up to unity at the present time.   The 
maximal-acceleration correction terms inside the integral were defined  
in ( 
4-8 )  :

$$\beta^2 =   F^2 ( z )  \equiv
[ H_o^2 R ( z ) L_c ] ^2  [   \Omega_\Lambda  - { \Omega_m \over 2 }  ( 
1 + 
z )^3   ] ^2  $$
with the requirement that :  $  0 \leq F^2 ( z )  \leq 1 $ , otherwise 
the 
square root would have been imaginary in ( 4-19)

The  temporal displacement  $ t ( z _o ) - t ( z_1)  >  0 $ is    :

$$ t ( z _o) - t (z_1 )  =   \int_{z_o} ^{ z_1}   { d z \over   H ( z ) 
}    
 > 0 . ~~~ z_1 > z_o         . \eqno ( 4.22 ) $$

The size $ R ( t )  \equiv a ( t ) R $  is recast in terms of the 
redshift  
:

$$ R  ( z )  = { R ( z_o ) \over 1 + z }  = { R_o \over  1 + z } . ~~~ 
z_o = 
0   $$

$$  {  ( dR ( t ) / dt )  \over R  } =  -  {  ( d z/dt )  \over 1 + z }  
=   
- { H ( z  ) \over  1 + z } $$
$$ \Rightarrow   { dR \over R }  = -  { d z \over 1 + z } . \eqno ( 
4.23 )  
$$

All the terms involved in the fundamental equation (  4-19  ) that  
furnish 
the fractional change of the fine structure constant  due to the 
maximal-acceleration relativistic corrections are now defined.  Had the 
maximal acceleration been
$A = \infty \leftrightarrow L_c= 0 $,  the equation ( 4-19 ) would have 
yielded
a $zero$ fractional change because in this extreme case  $ F^2 ( z ) = 
0 $,
for $all$  $ z$,  the  integrand of  eq-( 4-19) becomes unity
and one gets after performing the trivial time integral  :

$$  { \Delta \alpha \over \alpha } ( A = \infty ) =
[   ( t ( z_1  ) - t ( z_o ) )  / (  t ( z_1 ) - t ( z_o ) )  ]   - 1  
= 0 . 
\eqno ( 4.24) $$.
We shall study now the range of parameters in order for the fundamental 
equation to be well defined and furnish sound physical results 
compatible 
with observations.
The crux of this work is to set the cutoff  scale $  R ( t ) \ge  L_c   
\ge  
L_{Planck } $
from the requirement  that  close to the  cutoff $ L_c$  the 
maximal-acceleration corrections were the most dominant .   At the 
present 
epoch  $ z_o = 0 $ ,
$ R_o = R (  z_o ) = a_o R $ ,

The $  \beta^2 = F^2 (z) $ factors must obey   the  maximal 
acceleration 
relativistic constraints  $ 0 \leq  F^2(z) \leq 1 $ .  Upon using the 
definition of the Cosmological redshift :

$$  R ( z )  =   {  R_o \over 1 + z }    . ~~~
R_o =    R ( z_o )  = a_o R  . \eqno ( 4.25) $$

one gets    :

$$  0 \leq   [ H^2 _o R_o   L_c   ]^2  ( 1 +  z   )^{ - 2 }  [   
\Omega_\Lambda  -
{ \Omega_m \over 2 }  ( 1 + z  )^3   ]^2     \leq  1 . \eqno ( 4.26)  
$$

If one requires that the Maximal-acceleration relativistic effects are  
dominant at the cutoff scale $ L_c $ such that

$$ R_{min}  \equiv   R ( z_{max} ( L_c ) )  = L_c  \ge  L_P . \eqno ( 
4.27 a 
)  $$

the equation  which $defines$ the explicit relation between  $ z_{max } 
$ 
and $ L_c $  is obtained by demanding that the upper limiting 
acceleration  
at  $ L_c $  gives  :

$$ z = z_{max} ( L_c )  \Rightarrow  F^2 ( z_{max} )  = 1 \Rightarrow  
$$
$$ [  H^2_o  R_o  L_c  ]^2  ( 1 + z_{max} )^{ -2 }
[   \Omega_\Lambda  - { \Omega_m \over 2 }  ( 1 + z_{max}  )^3   ] ^2   
=   
1 . \eqno ( 4.27b) $$

The most interesting case is when we set  the minimum scale to coincide 
precisely with the Planck scale :
$ R_{min} = R ( z_{max} ) = L_c = L_P $  that implies that $ z_{max} 
\rightarrow \infty $ .
If one takes the  characteristic scale of the RWF metric $ R $  to 
coincide 
with the Hubble radius-horizon ( as observed today )  $ R = R_H  =  ( 
1/H_o 
) $, in units of $ c = 1 $,  to be $ 10^{60}-10^{61} $  Planck lengths 
$ = 
10^{27}-10^{28} cms$.
In this case it will be $meaningless$ to have parameter values like $ 
\Omega_m \sim 0.3$ for the simple reason that  $ \beta^2 =  F^2 ( z )  
>>> 1 
$ contrary to the maximal-acceleration relativistic principle that the 
upper 
bound on the acceleration cannot be surpassed , where $ A_{max} = c^2/ 
L_P $ 
  is the maximum in this case.  The $ \Omega_m$ in this extreme case 
has to 
be basically $ zero$ if we wish to  satisfy :

$$ L_P = L_c = R_{min}  = R(z_{max})  \Rightarrow $$
$$ F^2 ( z )   =  [ H_o  L_P ]^{ 2} [ H_o R_o ]^2   ( 1 + z_{max} )^{ 
-2 }
[ \Omega_\Lambda - {\Omega_m \over 2 }  ( 1 + z _{max} )^3 ]^2   \sim 1 
. 
\eqno ( 4.28 ) $$
For the choice  $ R = R_H  =  ( 1/H_o ) $  , the  curvature parameter  
as 
the scaling factor $ a_o \rightarrow \infty$  gives :
$$ \Omega_R  =   (   { 1 \over H_o a_o R }  )^2   =  ( { 1 \over a_o } 
)^2 
\rightarrow 0 . \eqno ( 4.29) $$
Hence  one concludes that as $ z_{max} \rightarrow  \infty$   eq-(4-28) 
becomes :
$$  [ H_o  L_P ]^{ 4}   ( {\Omega_m \over 2 } )^2   ( z _{max} )6    
\sim 1 
\Rightarrow \Omega_m \rightarrow  0 . \eqno ( 4.30 ) $$
otherwise it would have been impossible to counter-balance the huge 
factors 
stemming from the $ z_{max}^6 $  terms despite the  very small factor
$[ H_o  L_P ]^{ 4} \sim 10^{ - 240 } - 10^{ - 244 } $.

Therefore in this extreme case scenario because of  the condition :

$$ \Omega_\Lambda + \Omega_m \pm  \Omega_R = 1  . \eqno ( 4.31 )  $$
requires, due to  $ \Omega_R \rightarrow 0 $ and  $  \Omega_m 
\rightarrow 0 
$,  that  $ \Omega_\Lambda \rightarrow 1 $ ! ! !.

An important remark is in order.  One has to be very careful in 
distinguishing the  infinite acceleration case which gives $ F^2( z )  
= 0 
$, for $all$ values of $ z $ , and consequently the fractional change 
is  
identically $ ( \Delta \alpha / \alpha  ) = 0 $.
And the extreme case scenario $ F^2 ( z_{max} ) \sim 1 $ ,  for a  very 
particular value of the redshift  when the cutoff $ L_c = L_P \not= 0 $ 
,  
that yields a  very , very large value of $ z_{max}$ , but not $ 
\infty$ .  
In this case the fractional change  after computing the integrals  ( 
4-19 ) 
is $not $ zero, but in fact, it attains its limiting value in 
magnitude.

This extreme case $ \Omega_\Lambda = 1 $  is the most interesting 
because of 
its  implications :  Physically it suggests  that the universe was an 
extremely unstable
cosmic ? egg ? ( or ? atom ? )  of Planck size,  since there cannot be  
an 
initial point singularity as such due to the minimal Planck scale 
principle, 
  at  the  maximum attainable Planck temperature $ T_P $ ,  that 
exploded  
due to the huge vacuum instability, with an initial speed of $ c $ and  
maximal acceleration $ A = c^2/ L_P $ .  It was a true inflationary 
explosion driven by the enormous vacuum energy $ \Omega_\Lambda  \sim  
1 $ .
Eventually  the expansion began to slow down while the acceleration 
began to 
decrease,  from its maximal value  $ c^2/ L_P$ to the presently 
observed 
acceleration,  as a result of the ensuing attractive gravitational 
forces 
among the emerging material constituents of the Universe.  Matter was 
being 
created out of the vacuum to halt down the huge initial acceleration  $ 
A = 
c^2/ L_P $ .  In this respect this model is not very different than 
Hoyle?s 
steady state cosmology where matter was being created as the Universe 
expanded in order to maintain the matter density constant.

Closely related to this issue is the cosmological constant problem .
Within the framework of the Extended Scale Relativity Theory [ 1 ] this 
is 
an ill posed problem for the simple reason that in C-spaces the vacuum 
energy is just $one$ component of a Clifford-valued geometric object,  
a 
polyvector,  in the same way that the energy is just the component of a 
four-vector in ordinary Relativity.  Therefore, in C-spaces, the vacuum 
energy is $not$ a constant  but it changes under C-space Lorentz 
transformations ! . It was shown in [ 1 ] why the Conformal group 
originates 
from Clifford algebras, henceforth C-space Lorentz transformations 
contain 
$scalings$ which imply that the vacuum energy  itself is subjected to  
Renormalization-Group scaling-like flows as the Universe expands.  In 
C-spaces there are two times. The standard coordinate time, and the 
Stuckelberg-like time represented by the volume of the Universe, the 
cosmological clock, an arrow of time.

Within the framework of Nottale ?s Scale Relativity the cosmological 
problem 
is due to the fact that it is $meaningless$ to compare the vacuum 
energy at 
two separate scales that  differ in $ 60-61$ orders of magnitude, 
without 
taking into account scale relativistic corrections, like one does in 
ordinary Lorentz transformations.

Nottale has shown that the scale relativistic corrections must be such 
:

$$  { \Lambda ( Planck ) \over \Lambda ( R_H ) }    =
(   { R_H \over L_P }  )^2  \sim 10^{120}-10^{122}. \eqno ( 4.32 )  $$

This explains the orgins of such huge orders of magnitude discrepancy 
between the vacuum energy densities  at such different scales.

For a very intreresting application of this extreme case scenario, 
within 
the context of  scaling in cosmology , the arrow of time , the 
variation of 
the fundamental ? constants ? in Nature and the  plausible reason 
behind the 
Dirac-Eddington large number coincidences we refer to [ 14 ] .  The 
model in 
[ 14 ] is also based on the cosmic ? egg ? idea where the Hubble 
horizon is 
expanding  precisely with the speed of light which suggests that the 
Universe is the ultimate ? black hole ? .   If  matter truly is being 
created out of the vacuum, as the Hubble horizon radius grows,  its 
matter 
content grows accordingly in the same fashion that the Schwarzschild 
radius 
grows linearly with the  black hole Mass contents inside.  There are 
some 
important differences.
As we said above, the Extended Scale Relativity  theory has two times, 
the 
ordinary clock time and the scaling cosmological time representing the  
volume of the Universe. It precludes point singularities as such  due 
to the 
minimal scale principle, for this reason the Planck temperature is the 
maximum temperature in Nature ; i.e. Hawking evaporation $stops$  at 
the 
Planck scale when the black hole attains its limiting Planck 
temperature [ 1 
] .

From eq-(4-19) ,  when the maximal-acceleration effects are dominant at 
the
scale $ L_c$,  we can see why the integral term is less than unity and 
the 
fractional change of the fine structure constant is an increasing 
function  
of time ( decreasing function of $ z $ ) .  The  minus sign in
$ \Delta \alpha $  indicates that the electric charge $ e ( t ) $ in 
the 
past was lower than today  $ e ( t_o ) $ , and consequently the fine 
structure constant was lower in the past which is consistent with what 
has 
been observed [ 11 ] .

The present time orgin $ t_o $  is defined such as  whenever  one sets
$ t _1= t_o \Rightarrow  \Delta \alpha = 0  $  as required.
Since ,  when $ t_1 = t_o $, eq-( 4-19 ) becomes  when $ z_o = z_1 $ :

$$ {  \Delta \alpha \over \alpha }  =
[  (  {  \sqrt { 1 -   F ( z_o)^2 }  \over  \sqrt { 1 -   F ( z_o)^2 }  
}  ) 
( t_o - t_o ) / ( t_o - t_o )  ]
- 1 =  1 - 1 =   0 . \eqno ( 4.33 ) $$
as it should, by definition.

Collecting all these results we arrive finally at the fundamental 
equation 
that yields the cosmological-time variation of the fine structure 
constant 
in terms of the
cosmological redshift-factors  and the cutoff scale  $ L_c$  such that
$ R_o >  R ( z )  >   R ( z_{max}  )   =  R_{min}  =  L_c    \ge    
L_{Planck} $ :

$$  ( { \Delta ( \alpha )  \over   \alpha  } )   ( z;  L_c )  =    - 1 
+
   { 1 \over  ( t_o  - t ( z ) )    \sqrt { 1 - F^2 ( 0 ) }   }
\int _{0}^{z }   ~ { dz \over H(z) }  \sqrt  { 1 - F^2 ( z )  }  <  0 . 
\eqno ( 4.34 )  $$  .

The fractional change is an explicit function of the  running variable 
$ z $ 
  and the cutoff scale $ L_c$ exactly as it occurs in the Standard 
Renormalization Group methods in QFT.  This is not surprising in view 
of the 
ultraviolet/infrared entanglement which is the hallmark of 
Noncommutative  
Field theories based on Noncommutative Geometry at the Planck scale.  
The 
Planck scale $ L_{Planck} $  pays the role of the Noncommutative 
parameter 
of the spacetime coordinates.

$ T  = t_o - t ( z ) $ is the relative age of the Universe measured 
with 
respect to the  time given by  $ t (  z )  < t_o $ :

$$  t_o  -  t ( z )  =  \int_0^{  z }   { dz \over H ( z ) }. \eqno ( 
4.35 ) 
  $$
with  [ 13 ] :

$$H ( z ) \equiv  H_o ( 1 + z ) [ \Omega_m ( 1+z )^3 \pm  \Omega_R ( 1 
+ z )^2  + \Omega_\Lambda]^{ 1/2 } $$
Naturally when   $ z = z_o = 0 $ one gets $ t_o  - t_o = 0 $.
For  the following sequence of scale-orderings :

$$ 0    <   L_{Planck}     \leq     L_c      \leq     R ( z  )   <  R ( 
z_o 
)  = R_o $$

one has a lower and upper bound on the fractional change :

$$  -1 <   x_{Planck}    \leq  ( { \Delta \alpha \over \alpha} ) ( z,  
L_c ) 
  < 0 . \eqno ( 4.36 ) $$
meaning that there  is an upper bound in the $magnitude$  of the
fractional change  when  $  R ( z_{max} ) = R_{min} = L_c  = L_P $

The fundamental equation ( 4-34) is written in terms of $ all $ the 
fundamental cosmological parameters.   One can then use this 
fundamental 
equation for the  variation of the fine structure constant  to tune in 
precisely all the fundamental cosmological parameters

$$ H_o, ~\Omega_m ~ , \Omega_\Lambda~ , R_o , ....    . \eqno ( 4.37 ) 
$$
in terms of  the  cosmological redshift $ z $ and the cutoff scale $ 
L_c$ , 
and to check wether or not, one gets sound numerical results.  The fact 
that 
the maximal-acceleration corrections to the Lamb-shift [ 12 ] gave 
corrections to the fine structure constant of the order of $ 10^{ -5 } 
$ is 
very encouraging.  From this model we derived the fundamental equation 
above.

Variations of the fine structure constant have been observed over way 
much 
smaller redshifts than $ z_{max}$. Consequently the cutoff scale $ L_c$ 
must 
be taken to be larger than the Planck scale to match the present  
fractional 
change observations of  the order of  $ - 10^{ -5 } $ .  In this case, 
the 
cutoff scale $ L_c$  becomes  another cosmological parameter that must 
be 
tuned in accordingly
to yield sensible and verifiable predictions.

The main point of this work is that  if the maximal-acceleration had 
been $ 
infinity$,   then  $ \Delta \alpha = 0 $  identically  ! .   In the 
future,  
to be more rigorous and precise,  we shall use the full-fledged C-space 
Gravitational equations of motion which are tantamount to a Higher 
derivative Gravity with Torsion plus a C-space generalization of 
Yang-Mills 
and Electrodynamics [  1 ] .  i.e we must modify entirely the present 
Cosmological models in order to understand the Planck scale regime of 
the 
Universe. Current models based on conventional theories simply do not 
work.  
We require then to quantize the C-space Extended Relativistic Field 
theory  
using  tools based on Braided Hopf Quantum Clifford algebras, 
q-deformations 
of Clifford algebras,  for example.

In previous work we evaluated the running Planck constant due to the 
Extended Scale Relativistic Effects in C-spaces and which was 
responsible 
for the string/brane minimal length/time  uncertainty relations  [ 1 ]  
:

$$ \hbar_{eff} ( E )  = \hbar_o  { sinh ( L_P E ) \over \L_P E}. ~~~ 
\hbar_o = 
c = 1.
\eqno ( 4.38)  $$

when $ L_P \rightarrow 0 $ , or at very low energies ,
the $ \hbar_{eff}  =  \hbar_o $  as it should.
At Planck?s energy we have  $ E L_P \sim 1 $  :

$$ \hbar_{eff} ( Planck  ) = \hbar_o  sinh ( 1 ) . \eqno ( 4.39) $$
The fine structure constant is then :

$$  { e^2 \over  \hbar_{eff} ( Planck ) c }   <    { e^2 \over \hbar_o 
c } = 
{ 1 \over 137 } .
\eqno ( 4.40) $$
this means that the fine structure would have been smaller at lower 
scales,
for $fixed$ value of the Cosmological time = scaling size of the 
Universe.
To compensate for the running value of $ \hbar $  that would have 
induced a 
lower value of the fine structure constant,  for $fixed$ value of the 
cosmological clock, we must have a running value of the electric charge  
(  
the speed of light is unaltered ) :

$$  { e^2 _{eff} ( Planck ) \over  \hbar_{eff} ( Planck ) c } =
{ 1 \over 137 } { 1 \over sinh ( 1 ) }  { e^2_{eff } ( Planck ) \over 
e^2 }  
=
{ 1 \over 4 \pi^2 } . \eqno ( 4.41) $$
where we used the Nottale  value for the fine structure constant
at the Planck scale  of $ { 1 / 4 \pi^2 } $
From eq- ( 4-41 ) we get  a fractional increase of the electric charge 
;

$$ { e^2_{eff } ( Planck ) \over e^2 ( m_e) } =  { 137 \over 4 \pi^2 }  
sinh 
( 1 )  > 1.
\eqno ( 4.42 )  $$
which is indeed consistent with the standard Renormalization Group 
arguments
in { \bf QED} with the extra modification of the $ sinh ( 1 ) $ factor.

Concluding   :   The fine structure constant can vary either :

$ \bullet$   by changing the value of the Cosmological clock = scale 
size of 
the Universe and , as we have shown ,  its value was $lower$ in the 
past 
than it is today due to the Maximal-Acceleration Relativistic effects .  
Had 
the maximal acceleration been infinity, the $ \Delta \alpha / \alpha = 
0 $ !
The maximal-acceleration corrections to the Lamb-shift were calculated 
by [ 
12 ] and yield fractional changes to the fine structure constant of  
the 
order of $ 10^{-5} $.

or

$\bullet $ by  increasing  the energy  (  probing smaller distances )  
from 
the electron?s Compton wavelength to the Planck?s scale,  keeping fixed 
the 
Cosmological clock.

We believe that these two last points are highly nontrivial and may 
reveal 
new Physics in the horizon.  In particular, we must modify the current 
Quantum Field Theories to be able to  incoporate the 
maximal-acceleration 
relativistic effects .  Beginning,
for example, by studying the appropriate Noncommutative Quantum 
Mechanics  
due to the Extended Scale Relativistic dynamics in C-spaces ( Clifford 
manifolds )  [ 1 ] and the canonical groups ( with their 
representations ) 
acting on Noncommutative extended phase spaces [ 16 ] .

\centerline{\bf  Acknowledgements}

We are kindly indebted to M.Bowers and J. Mahecha for their assistance 
in
preparing the manuscript and hospitality in Santa Barbara where this 
work 
was completed.  We thank S. Low for sending us reference [ 16 ] .

\bigskip

\centerline { \bf References }

\bigskip

1 - C. Castro, ``The programs of the Extended Relativity in C-spaces, 
towards
the physical foundations of String theory", hep-th/0205065. To appear 
in the
proceedings of the NATO advanced workshop on the nature of time, 
geometry,
physics and perception. Tatranska Lomnica, Slovakia, May 2002.  Kluwer
Academic Publishers;

``Noncommutative Quantum Mechanics and Geometry from the quantization 
of C-
spaces", hep-th/0206181.

? Maximal-Acceleration Phase Space Relativity from Clifford Algebras ? 
hep-th/0208138

C. Castro, M. Pavsic : ? Higher Deriavtive Gravity and Torsion from the 
Geometry of C-spaces ? Phys. Lett { \bf 539 } ( 2002 ) 133.  
hep-th/0110079.
? The Clifford Algebra of spacetime and the conformal group ? 
hep-th/0203194.

2 - L. Nottale, ``Fractal Spacetime and Microphysics, towards Scale
Relativity", World Scientific, Singapore, 1992;
``La Relativite dans tous ses etats", Hachette Literature Paris, 1999.

3 - M. Pavsic, ``The landscape of Theoretical Physics: A global view  
from
point particles to the brane world and beyond, in search of a unifying
principle",  Kluwer Academic Publishers 119, 2001.

4 - E. Caianiello, ``Is there a maximal acceleration?", Lett. Nuovo 
Cimento
{\bf 32} (1981) 65.

5 - V. Nesterenko, Class. Quant. Grav. {\bf 9} (1992) 1101; Phys. Lett. 
{\bf
B 327} (1994) 50;
V. Nesterenko,  A. Feoli, G. Scarpetta, ``Dynamics of relativistic part

6-H. Brandt :  Contemporary Mathematics {\bf 196 } ( 1996 ) 273.
Chaos, Solitons and Fractals {\bf 10 } ( 2-3 ) ( 1999 ) .

7-M.Pavsic : Phys. Lett {\bf B 205 } ( 1988) 231 ; Phys. Lett {\bf B 
221} ( 
1989 ) 264.
H. Arodz , A. Sitarz, P. Wegrzyn : Acta Physica Polonica { \bf B 20} ( 
1989 
) 921.

8- F. Schuller :   ? Born-Infeld Kinematics and corrections to the 
Thomas 
precession ?
hep-th/0207047, Annals of Phys. {\bf 299 } ( 2002 ) 174.

9- A. Chernitskii : ? Born-Infeld electrodynamics, Clifford numbers and 
spinor representations ? hep-th/0009121.

10- J. Lukierski, A. Nowicki, H. Ruegg, V. Tolstoy : Phys. Lett { \bf 
264 }  
( 1991 ) 331.
J. Lukierski, H. Ruegg , W. Zakrzewski : Ann. Phys { \bf 243 } { 1995 ) 
90.

G.Amelino-Camelia :  Phys. Lett { \bf B 510} ( 2001 ) 255.
Int. J. Mod. Phys { \bf D 11 } ( 2002 ) 35, gr-qc/0012051

S. Rama : ? Classical velocity in kappa-deformed Poincare algebra and a 
Maximal Acceleration ? hep-th/0209129.

11- J. Webb, M. Murphy, V. Flambaum, V. Dzuba,  J. Barrow, C. 
Churchill, J. 
Prochaska, and A. Wolfe : ?  Further evidence for Cosmological 
Evolution of 
the Fine Structure Constant ?  astro-ph/0012539.

12- G. Lambiase, G.Papini. G. Scarpetta :  ? Maximal Acceleration 
Corrections to the Lamb Shift of one Electron Atoms ? hep-th/9702130.

13- H. Peebles  : ? Principles of Physical Cosmology ? . Princeton 
University Press.

J. V. Narlikar, T. Padmanabhan : ? ? Gravity, Gauge Theories and 
Quantum 
Cosmology ? Kluwer Academic Publishers ,  1986.

14- M. Kafatos, S. Roy, R. Amoroso :  ? Scaling in Cosmology and the 
arrow 
of time ?
Studies on the Structure of Time . Edited by Buccheri et al. Kluwer 
Academic. Plenum Publishers , New York , 2000,  191-200.

J. Glanz  : Science  {\bf 282 } ( 1998 ) 2156.

15- L. Castellani : Phys. Lett {\bf  B 327 } ( 1994 ) 22.  Comm. Math. 
Phys 
{ \bf 171 }
( 1995 )  383.

16- S. Low :  Jour. Phys { \bf A  } Math. Gen {\bf 35 }  ( 2002 ) 5711.

\end{document}